\documentclass[reprint, amsmath, amssymb, aps, prx, superscriptaddress, floatfix, nofootinbib, citeautoscript, longbibliography]{revtex4-1}

\usepackage[utf8]{inputenc}
\usepackage{graphicx}
\usepackage{dcolumn}
\usepackage{bm}
\usepackage{braket}
\usepackage{color}
\usepackage{units}
\usepackage{dsfont}
\usepackage{float}
\usepackage{hyperref}
\usepackage{xcolor}

\begin{document}
\setlength{\abovedisplayskip}{0pt}
\setlength{\belowdisplayskip}{6pt}
\abovedisplayshortskip=0pt
\belowdisplayshortskip=6pt
 
\title{Overcoming Noise in Entanglement Distribution}

\author{Sebastian Ecker}
\email{sebastian.ecker@oeaw.ac.at}
\affiliation{Institute for Quantum Optics and Quantum Information (IQOQI), Austrian Academy of Sciences, Boltzmanngasse 3, 1090 Vienna, Austria.}
\affiliation{Vienna Center for Quantum Science and Technology (VCQ), Faculty of Physics, University of Vienna, Boltzmanngasse 5, 1090 Vienna, Austria}

\author{Fr\'{e}d\'{e}ric Bouchard} 
\affiliation{Department of physics, University of Ottawa, Advanced Research Complex, 25 Templeton, Ottawa ON Canada, K1N 6N5}

\author{Lukas Bulla}
\affiliation{Institute for Quantum Optics and Quantum Information (IQOQI), Austrian Academy of Sciences, Boltzmanngasse 3, 1090 Vienna, Austria.}
\affiliation{Vienna Center for Quantum Science and Technology (VCQ), Faculty of Physics, University of Vienna, Boltzmanngasse 5, 1090 Vienna, Austria}

\author{Florian Brandt}
\affiliation{Institute for Quantum Optics and Quantum Information (IQOQI), Austrian Academy of Sciences, Boltzmanngasse 3, 1090 Vienna, Austria.}
\affiliation{Vienna Center for Quantum Science and Technology (VCQ), Faculty of Physics, University of Vienna, Boltzmanngasse 5, 1090 Vienna, Austria}

\author{Oskar Kohout}
\affiliation{Institute for Quantum Optics and Quantum Information (IQOQI), Austrian Academy of Sciences, Boltzmanngasse 3, 1090 Vienna, Austria.}
\affiliation{Vienna Center for Quantum Science and Technology (VCQ), Faculty of Physics, University of Vienna, Boltzmanngasse 5, 1090 Vienna, Austria}

\author{Fabian Steinlechner}
\affiliation{Institute for Quantum Optics and Quantum Information (IQOQI), Austrian Academy of Sciences, Boltzmanngasse 3, 1090 Vienna, Austria.}
\affiliation{Vienna Center for Quantum Science and Technology (VCQ), Faculty of Physics, University of Vienna, Boltzmanngasse 5, 1090 Vienna, Austria}
\affiliation{Fraunhofer Institute for Applied Optics and Precision Engineering IOF, Albert-Einstein-Strasse 7, 07745 Jena, Germany}
\affiliation{Abbe Center of Photonics - Friedrich-Schiller-University Jena, Albert-Einstein-Str. 6, 07745 Jena, Germany}

\author{Robert Fickler}
\email{robert.fickler@tuni.fi}
\affiliation{Institute for Quantum Optics and Quantum Information (IQOQI), Austrian Academy of Sciences, Boltzmanngasse 3, 1090 Vienna, Austria.}
\affiliation{Vienna Center for Quantum Science and Technology (VCQ), Faculty of Physics, University of Vienna, Boltzmanngasse 5, 1090 Vienna, Austria}
\affiliation{Photonics Laboratory, Physics Unit, Tampere University, Tampere, FI-33720, Finland}

\author{Mehul Malik}
\email{m.malik@hw.ac.uk}
\affiliation{Institute for Quantum Optics and Quantum Information (IQOQI), Austrian Academy of Sciences, Boltzmanngasse 3, 1090 Vienna, Austria.}
\affiliation{Vienna Center for Quantum Science and Technology (VCQ), Faculty of Physics, University of Vienna, Boltzmanngasse 5, 1090 Vienna, Austria}
\affiliation{Institute of Photonic and Quantum Sciences (IPaQS), Heriot-Watt University, Edinburgh, Scotland, UK EH14 4AS}

\author{Yelena Guryanova}
\email{yelena.guryanova@oeaw.ac.at}
\affiliation{Institute for Quantum Optics and Quantum Information (IQOQI), Austrian Academy of Sciences, Boltzmanngasse 3, 1090 Vienna, Austria.}
\affiliation{Vienna Center for Quantum Science and Technology (VCQ), Faculty of Physics, University of Vienna, Boltzmanngasse 5, 1090 Vienna, Austria}

\author{Rupert Ursin}
\email{rupert.ursin@oeaw.ac.at}
\affiliation{Institute for Quantum Optics and Quantum Information (IQOQI), Austrian Academy of Sciences, Boltzmanngasse 3, 1090 Vienna, Austria.}
\affiliation{Vienna Center for Quantum Science and Technology (VCQ), Faculty of Physics, University of Vienna, Boltzmanngasse 5, 1090 Vienna, Austria}

\author{Marcus Huber}
\email{marcus.huber@univie.ac.at}
\affiliation{Institute for Quantum Optics and Quantum Information (IQOQI), Austrian Academy of Sciences, Boltzmanngasse 3, 1090 Vienna, Austria.}
\affiliation{Vienna Center for Quantum Science and Technology (VCQ), Faculty of Physics, University of Vienna, Boltzmanngasse 5, 1090 Vienna, Austria}

\begin{abstract}
\textbf{Noise can be considered the natural enemy of quantum information. An often implied benefit of high-dimensional entanglement is its increased resilience to noise. However, manifesting this potential in an experimentally meaningful fashion is challenging and has never been done before. In infinite dimensional spaces, discretisation is inevitable and renders the effective dimension of quantum states a tunable parameter. Owing to advances in experimental techniques and theoretical tools, we demonstrate an increased resistance to noise by identifying two pathways to exploit high-dimensional entangled states. Our study is based on two separate experiments utilising canonical spatio-temporal properties of entangled photon pairs. Following these different pathways to noise resilience, we are able to certify entanglement in the photonic orbital-angular-momentum and energy-time degrees of freedom up to noise conditions corresponding to a noise fraction of 72~\% and 92~\% respectively. Our work paves the way towards practical quantum communication systems that are able to surpass current noise and distance limitations, while not compromising on potential device-independence.}
\end{abstract}

\maketitle

\section{Introduction}
Quantum entanglement is one of the most peculiar and elusive properties of quantum systems, a key resource in quantum information processing \cite{Friis2019} and an indispensable ingredient for device-independent quantum cryptography \cite{Acin:2007db}. 
At the same time, entangled quantum systems are highly delicate since their entanglement is readily diminished by the slightest interaction with the environment. 
This is of particular relevance for the distribution of entangled photons over long distances outside of a protected laboratory environment, where particle loss and environmental noise are inevitable.
Similar to classical communication, noise ultimately reduces the channel capacity and thus acts as a limiting factor for the link distance in quantum communications. 
Several proof-of-concept experiments have pushed the distribution distance of two-dimensional-entangled photon pairs over fiber  \cite{Salart2008,Inagaki2013,Wengerowsky2019} and free-space  \cite{Scheidl2009,Krenn2015,Yin:2017} links, while others have demonstrated the distribution of high-dimensional entangled quantum states \cite{Steinlechner:2017,Ikuta:2018,Cozzolino2019,liu2019multidimensional,Cozzolino2019air-core,cao2018distribution}.

Although it is not straightforward to certify high-dimensional entanglement from experimental data, its production in the process of spontaneous parametric down-conversion (SPDC) happens naturally.
As a result of  conservation laws in this process, the down-converted photon pairs are entangled in  spatio-temporal properties such as energy-time \cite{Thew2004, Brendel1999, Jha2008, Maclean2018}, angle-angular momentum \cite{Vaziri2002,Leach2010,Krenn2017,Erhard2018} and position-momentum \cite{Howell2004,Schaeff2015,Wang2018}. 

At first glance, from an abstract information theoretic point of view, high-dimensional entanglement might seem to be essentially reproducible by just many copies of regular qubit entanglement. While there is actually a notable difference even in idealised pure states \cite{Kraft2018} and cryptographic settings \cite{Huber2013}, one of the main reasons for developing high-dimensional protocols has predominantly been the aforementioned free availability in down-conversion combined with the capability of storing more bits per communicated photon. Indeed, many such benefits of using high-dimensional encodings in quantum key distribution (QKD) have been investigated in the last decade \cite{Bechmann-Pasquinucci:2000a,Bechmann-Pasquinucci:2000b,Cerf2002,Nikolopoulos:2005,Sheridan:2010}, followed by experimental implementations in recent years \cite{Ali-Khan:2007,Zhang:2013,Zhong:2015,Sit:17,Bouchard2018}.
Apart from an increased per-photon information capacity, an often implied advantage of employing high-dimensional entanglement is its potential for increased resistance to noise.

While it is indeed true that dimension-independent noise models show an increased resistance of entangled states to noise \cite{Zhu:2019tb,Collins2002}, the actual advantages very much depend on the physical implementation. Different high-dimensional degrees of freedom (DOFs) are bounded by different operational constraints. Thus, it has remained an open question whether practical improvements using high-dimensional entanglement can actually manifest its promised advantages. 

In this letter, we expound potential pathways to an increased resilience to noise by utilising entanglement in high dimensions. We conduct two experiments, exploiting the most paradigmatic platforms for generating high-dimensional entangled quantum states, namely photons entangled in energy-time as well as transverse position-momentum. We show that for each high-dimensional encoding method and its associated state-of-the-art technology, there is an appropriate pathway to verify entanglement in conditions where qubit entanglement cannot be distributed due to extreme external noise levels. We are further able to characterise a realistic trade-off between dimensionality and robustness to find optimal and flexible encodings for both implementations and different background conditions, thereby revealing the transformative potential of high-dimensional quantum information.

\begin{figure*}[t!]
\centering
\includegraphics[width=1\textwidth]{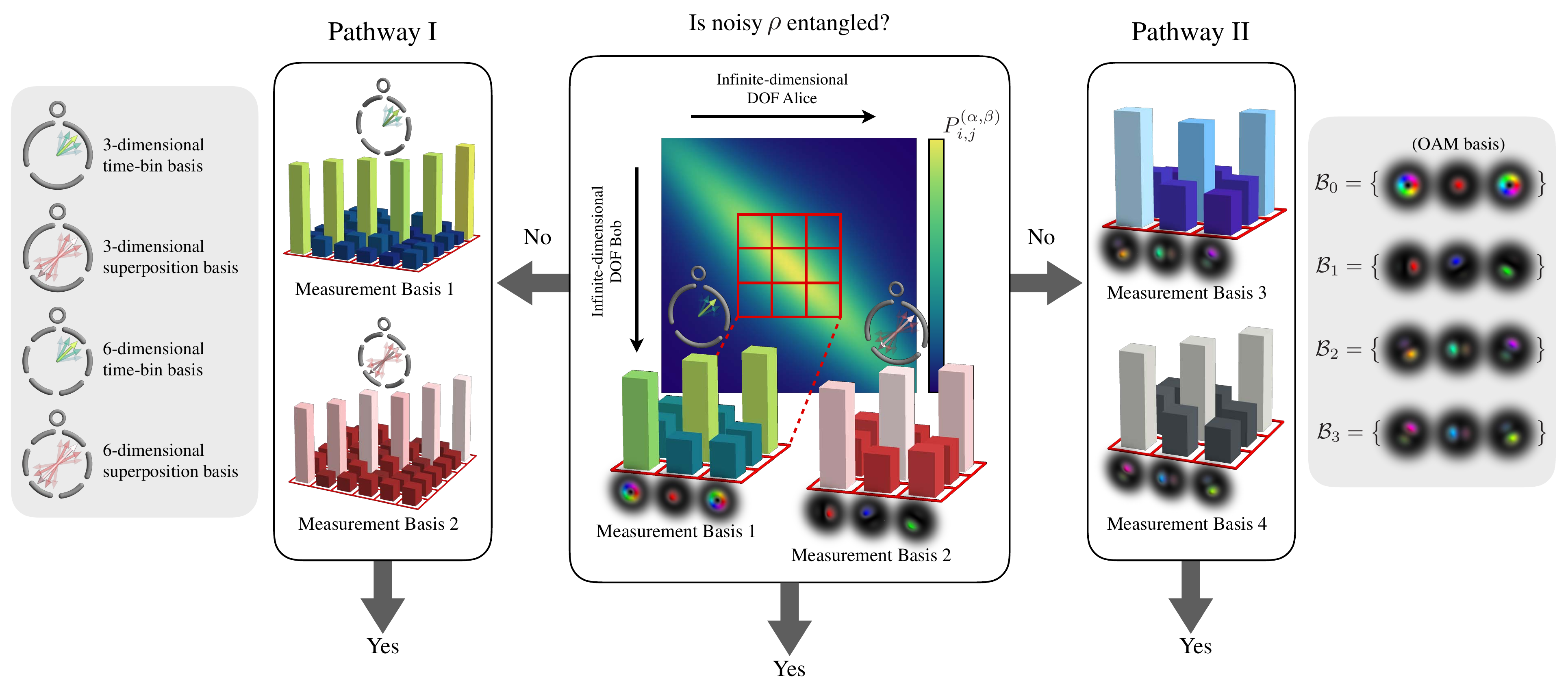}
\caption{Illustration of the pathways to noise resilience. A mixed entangled state $\rho$ shared by Alice and Bob is encoded in spatio-temporal properties of photon pairs. Each of the infinite-dimensional degrees of freedom (DOF) of the photons can be discretised and measured in two bases (central panel). If both measurements are insufficient to certify entanglement in the noisy state, there are two pathways to recover it: Fine-graining to higher dimensions (Pathway I, left panel) and measuring in additional bases (Pathway II, right panel).  In Pathway I, noise is `diluted' by discretising the existing state space further, resulting in an increased signal-to-noise ratio. Pathway II exploits the existence of more than two mutually unbiased measurement bases in higher dimensions, providing additional information about the non-classicality of the state. The bar charts illustrate the joint probability $P_{i,j}^{(\alpha,\beta)}$ of measuring Alice's modes $i$ in the basis $\alpha$ and Bob's modes $j$ in the basis $\beta$.}
\label{fig:pathways}
\end{figure*}

\section{Pathways to noise resilience}

Almost all quantum experiments aim to harness a physical process that is expected to yield a pure entangled state. If the system is bipartite, and assuming that the experiment is ideal, then the entangled state can be represented in the Schmidt basis  $|\psi_{AB}\rangle=\sum_i\lambda_i|ii\rangle$. 
Needless to say, experiments are seldom ideal, and a number of factors contribute to spoiling the state, during both its generation and its manipulation. Errors could, for example, be introduced during the distribution of the state via quantum channels or through imperfect measurement devices. Moreover, background photons inevitably introduce noise, resulting in a reduction of the signal-to-noise ratio at the read-out. It is well known that noise deteriorates entanglement and the extent to which entanglement persists despite the presence of noise is known as `noise resistance of entanglement' \cite{Friis2019}.
The degree to which the initially pure state is degraded is often estimated using a white noise model, i.e. by mixing the target state $\ket{\psi}$ with the maximally mixed state:

\begin{equation}\label{eq:noisemodel}
\hat{\rho}=p \ket{\psi}\bra{\psi} + \frac{1-p}{d^2} \mathds{1}_{d^2}\,.
\end{equation}

One may also note that this model captures particle loss for the maximally entangled state $\ket{\psi}=\ket{\Phi^+}:=\frac{1}{\sqrt{d}}\sum_i|ii\rangle$, where with probability $p$ the state remains intact, and with probability $1-p$ a particle from a pair is lost. The measurement statistics of the lost photon correspond to the maximally mixed state, while the statistics of the partner photon are replaced by the marginal. In the case of the maximally entangled state, this marginal is also maximally mixed  $\text{Tr}_B\ket{\Phi^+}\bra{\Phi^+}=\frac{1}{{d}}\mathds{1}$, resulting in the model in~\eqref{eq:noisemodel}. For this `isotropic' state, the resulting tolerance to noise, i.e. the critical $p_{c}$ after which the state becomes separable, scales as $p_{c}=\frac{1}{d+1}$. This can already be concluded from the first criteria for mixed state entanglement, such as positivity under partial transposition \cite{PhysRevLett.77.1413,HORODECKI19961} and has already been pointed out in early literature \cite{Vidal1999,Horodecki1999}. While for general states such resistance to depolarizing noise is quite generic \cite{Lami2016}, physical modeling can reveal even further avenues of avoiding noise in high-dimensions \cite{Zhu:2019tb}. 
We argue in Sec.~\ref{sec:expimpl}, that the noise introduced in both our experiments is close to white. Nonetheless, it is important to emphasise here, that we do not assume any noise model when analysing the experimental data for entanglement -- the simple noise model only serves as a motivating example for why we should be expecting an increased noise resistance and it is not needed for performing or analysing the experiment. In a realistic experimental setting, loss can affect the measurement statistics in more complex ways, such as introducing accidental coincidence counts due to detector or background noise. A more quantitative analysis of the precise role of noise in photonic entanglement has been performed recently \cite{Zhu:2019tb}, and supports our experimental results by demonstrating a clear advantage of going to high dimensions. For more general states, bipartite depolarizing maps \cite{Lami2016} capture different loss rates or detection efficiencies and can be solved analytically for any dimension. The common feature of these noise models is the fact that it is possible for the noise resistance to increase linearly with the system dimension $d_S$. As $d_S$ grows, so does the so-called `dimensionality of entanglement'. Thus, one should, in principle, be able to overcome any amount of noise, and detect entanglement, simply by looking in systems of high-enough dimension. Despite this feature, there are several reasons why this idea has not yet manifested in any practical setups. First, certifying entanglement requires one to collect enough information about the underlying quantum state. The number of measurements required to do this scales \textit{at least} linearly with the size of the system \cite{Bavaresco2018}. Second, the dimension of a system is not a fundamentally tunable parameter in an experiment and finally, the noise very much depends on the physical implementation of the chosen scheme. 
The theoretical description of spatio-temporal degrees of freedom of any photon state is infinite-dimensional,

\begin{align}
    \ket{\psi}_{\text{photon}} = \sum_s \int \text{d}\mu(x) \psi_s(x)\ket{x, s}
\end{align}

where $s$ is the polarisation DOF, $x$ the position and $\mu(x)$ a measure over the space. 
It then follows that the description of temporally or spatially entangled photon states is also infinite-dimensional. Despite this, any laboratory measurement still requires one to discretise these DOFs. The discretisation depends on the measuring device; for example, to discretise temporally entangled states one could time-resolve photon detections using high-precision clocks. For states entangled spatially, one could perform spatial mode measurements using spatial light modulators or cameras. All of these techniques have limited resolution; this means that by increasing the dimension of the states (i.e. discretising further) one will often encounter additional sources of noise, e.g. through cross-talk or additional measurement channels, which consequently lead to dimension-dependent noise factors $p(d)$ entering into the models. Thus, while high-dimensional entanglement presents an increased resistance to noise with increasing dimension on paper, it is not clear whether this theoretical advantage can be exploited in a real experiment. 

Nevertheless, noise resistance of entanglement is a highly desirable feature in quantum communication and is of utmost importance for fundamental reasons. If one is able to demonstrate the persistence of entanglement, simply by discretising the description of systems, then one may be closer to understanding the fundamental limits on the information capacity of single photon quantum communication channels. In spite of this potential, not a single quantum experiment to date has been able to show an increase in noise resistance in a controlled fashion. In this work, we present two experiments that discretise continuous DOFs to encode information in high-dimensional quantum systems to explicitly demonstrate an increased resilience to environmental noise. These paradigms are illustrated in Fig.~\ref{fig:pathways}. It is to be read as a flow chart, starting in the center where a hypothetical noisy quantum state $\rho$ is tested for entanglement by making measurements in two bases. If none is found, one has two options, depending on the DOF and its technological constraints. The first pathway (Fig.~\ref{fig:pathways} left) is to fine-grain or partition the quantum state to higher dimensions, for example by discretising an energy-time entangled state to a higher temporal resolution. Alternatively, one may explore pathway II  (Fig.~\ref{fig:pathways} right), which exploits the existence of more than two mutually unbiased bases (MUBs) in high dimensions. For example, measurements of an orbital-angular-momentum (OAM) entangled state can be made in additional bases, providing more information about the state. In both cases, entanglement can be recovered from a state in an assumption-free manner, where no entanglement could previously be certified through standard techniques.

\section{Experimental Implementation}
\label{sec:expimpl}

\begin{figure*}
\centering
\includegraphics[width=1\textwidth]{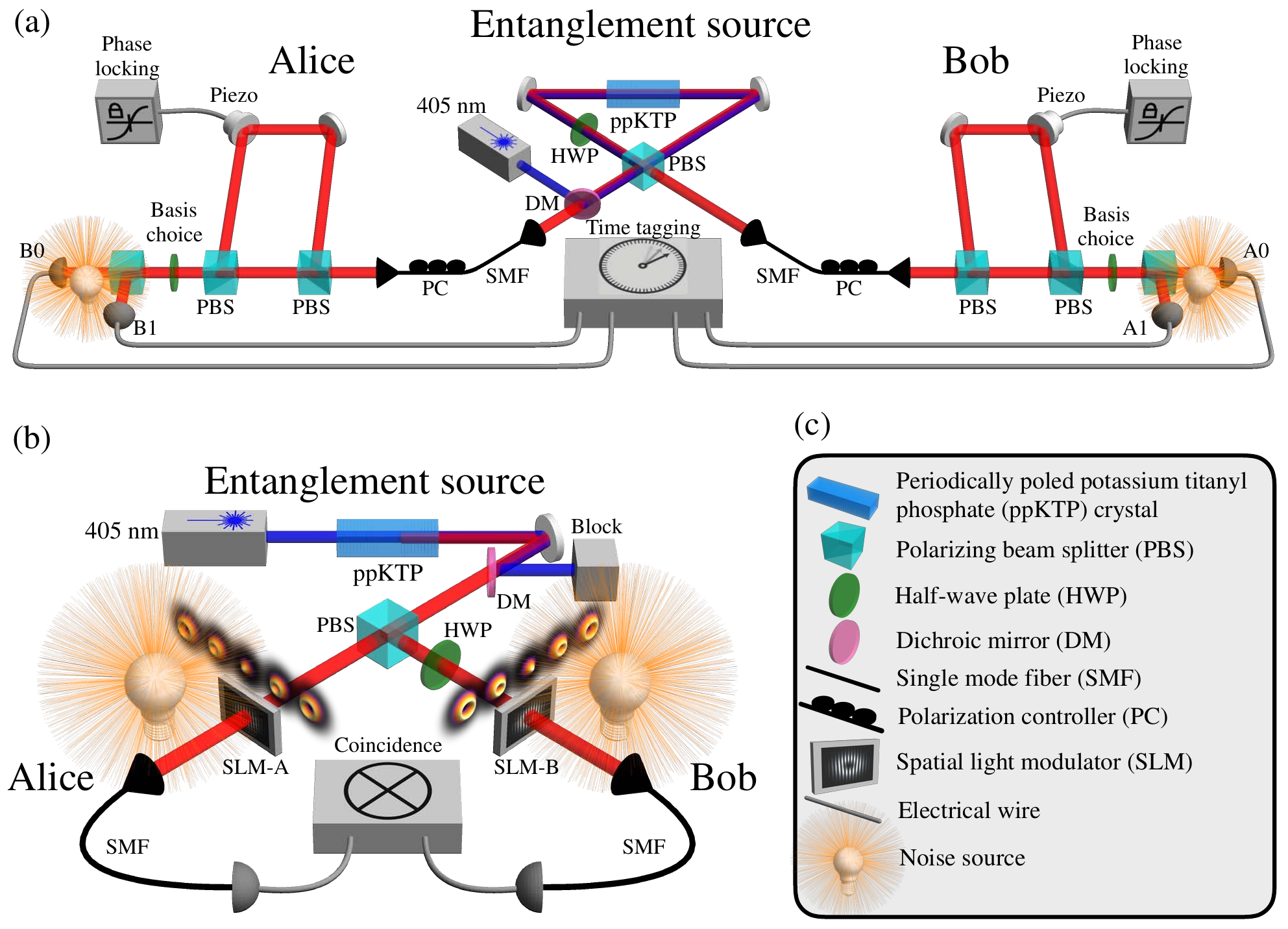}
\caption{Experimental setup for (a) energy-time and (b) orbital angular momentum (OAM) degrees of freedom. In both experiments, a  $\unit[405]{nm}$ continuous-wave laser produces high-dimensionally entangled photon pairs in a ppKTP crystal exploiting type-II spontaneous parametric down-conversion (SPDC). The noise is optically added by intensity-adjustable light sources and single-photon detection is accomplished using avalanche photo diodes. (a) Additional polarization entanglement is generated by bidirectionally pumping the crystal in a polarization Sagnac interferometer. The polarization basis the photon pairs are measured in after an  actively-stabilised post-selection-free Franson interferometer defines the measurement basis in the time domain. Each detection event is time-tagged and recorded by means of a time-to-amplitude converter. (b) The OAM-entangled pairs are split depending on their polarization, analysed through mode filtering by modulating the complex amplitude of the photons and subsequently coupled into SMFs. Coincidence counts are recorded using a coincidence logic.}
\label{fig:exp_setup}
\end{figure*}
Here, we showcase two photonic experiments that demonstrate high-dimensional noise resilience of entanglement via the above-described two pathways. In the first experiment we follow pathway I and exploit energy-time entanglement, while in the second experiment we take pathway II to explore the orbital-angular-momentum DOF, both encodings that have seen rapid experimental progress in recent years \cite{Friis2019}. The basic premise of both experiments is to create photon pairs, and show, via a set of appropriately chosen measurements, that these pairs remain entangled even in the presence of high levels of noise.
To generate the pairs in both experiments we appeal to spontaneous parametric downconversion (SPDC). 

First, let us consider the creation of photon pairs entangled in energy-time. In the nonlinear SPDC process, a crystal pumped with photons of frequency $\omega_P$ will spontaneously produce a pair of photons with frequencies $\omega_0$ and $\omega_1$. The total energy is strictly conserved such that, despite the crystal producing photon pairs with a finite bandwidth, the sum of their frequencies is constant: $\hbar \omega_P = \hbar \omega_0 + \hbar \omega_1 $. This results in the emission of two photons that are highly entangled in energy. Since the spectral linewidth and the coherence time are inversely related, a narrow pump bandwidth results in a long coherence time for possible photon pair emissions, giving rise to entanglement in the time-domain with Schmidt numbers up to $\sim10^9$ under realistic experimental assumptions \cite{Brougham2012,Ali-Khan:2007}.
In our scheme we utilize ancillary entanglement in the polarization DOF to facilitate interference in the time domain. 

A similar narrative holds for the second experiment, which produces photons entangled in the orbital-angular-momentum (OAM) DOF. Here, the strict conservation of momentum in the SPDC process $\hbar l_p = \hbar l_0 + \hbar l_1 $ results in the production of photon pairs anti-correlated in OAM $\hbar l_0 = -\hbar l_1$ for a Gaussian-mode pump photon with $\hbar l_p=0$, leading to entanglement in the OAM-angular position variables \cite{Krenn2017}. The (theoretically) infinite-dimensional states produced by the two experiments can be written as

\begin{align}
\ket{\Psi}_{\textrm{ET-pol}} &= \int dt f(t) \ket{t}_A\ket{t}_B \otimes \ket{\phi^-}_{AB}\\
\ket{\Psi}_{\textrm{OAM}}&=\sum_{\ell=-\infty}^{\infty} c_{\ell}\ket{-\ell}_A\ket{\ell}_B,
\end{align}
where $f(t)$ is a continuous function of time, corresponding to the coherence profile of the laser;  $\ket{\phi^-}_{AB} =\tfrac{1}{\sqrt{2}}(\ket{H}_A\ket{H}_B-\ket{V}_A\ket{V}_B)$ is a polarisation-entangled Bell state; $\ket{\pm\ell}$ is the state of a photon carrying an OAM quantum number of $\pm\ell$ and $c_\ell$ is a complex probability amplitude, which is defined by the spatial characteristics of the crystal and pump beam. 

In order to gain meaningful insight into noise resilience, both states must be appropriately discretised. In the energy-time experiment, we measure the time of arrival of entangled  photon pairs by discretising a time-frame of duration $F$ into bins and recording which bin a photon is detected in. The duration of $F$ is fixed and we divide it into an integer number of time-bin modes $d$, each corresponding to a duration $t_d$, i.e. $F/d =t_d$ (see Supplemental Material, Sec.~3 \cite{supplmat}). In pathway II, we choose a finite cut-off to the theoretically infinite sum over modes, such that the modes with OAM quantum numbers $l\in \{-D,...,D\}$ are spanning a $2D+1$ - dimensional Hilbert space. Thus, ideally, the states generated by the experiments would be close to the forms

\begin{align}\label{eq:etstate}
\ket{\Psi}_{\textrm{ET-pol}}&=\sum_{j=1}^{d} \alpha_{j}\ket{j}_A\ket{j}_B\otimes\ket{\phi^-}_{AB}\\
\ket{\Psi}_{\textrm{OAM}}&=\sum_{\ell=-D}^{D} c_{\ell}\ket{-\ell}_A\ket{\ell}_B, 
\label{eq:oamstate}
\end{align}
where $\ket{j}$ refers to a photon in a discrete time-bin state whose duration is $t_d$ for $j\in \{1, \ldots d \}$ and $\alpha_j$ is a complex probability amplitude. 

Despite investigating different DOFs, the experiments have similar characteristic features, as shown in Fig.~\ref{fig:exp_setup}. In both schemes, a nonlinear crystal is pumped with a continuous-wave diode laser to generate photon pairs, which then pass through a setup consisting of measurement elements and an external noise source. 
In addition, the entanglement dimensionality for both cases ($d$ for energy-time and $d=2D+1$ for OAM) is strongly dependent on the pump characteristics. In the energy-time experiment (Fig.~\hyperref[fig:exp_setup]{\ref*{fig:exp_setup}(a)}), a narrow-bandwidth pump ensures a large Schmidt number, while in the OAM experiment (Fig.~\hyperref[fig:exp_setup]{\ref*{fig:exp_setup}(b)}), a large pump mode with a well-defined transverse momentum results in high-dimensional OAM-entanglement. For additional experimental details, please see the Appendixes \hyperref[sec:appA]{A} and \hyperref[sec:appB]{B}. 
In both experiments, noise is introduced in the form of background photons generated by sources of light simulating a realistic operational environment for a quantum communication system. In the energy-time experiment, fine-adjustable light emitting diodes placed near the detectors introduce background counts---simulating a scenario where classical light may be co-propagating with a quantum signal. In the OAM experiment, background counts are introduced by increasing the intensity of the ambient light in the lab up to daylight conditions, which is a realistic scenario for free-space experiments using large aperture telescopes. White noise is generated in the energy-time experiment by employing two independent noise sources for Alice and Bob, thus eliminating temporal correlations, while in the OAM experiment white noise is introduced by placing the noise source after the spatial light modulators, ensuring mode-independent noise generation. In both cases, we quantify the amount of noise introduced via the noise fraction $N\!F$, which corresponds to the fraction of counts in our data that arise from noise. Intuitively, $N\!F = \frac{\#\text{noise counts}}{\#\text{total counts}}$, 
which takes on values from 0 (no noise) to 1 (complete noise). A more rigorous definition of the noise fraction $N\!F$ and its computation from experimental data is presented in the Supplemental Material, Sec.~2 \cite{supplmat}.

\section{Energy-time entanglement \\(Pathway I)}

 The first pathway to noise-resilience is implemented by fine-graining measurements of the photon arrival time. As outlined, we discretise a time-frame into $d$ time-bins and record the bin that a photon is detected in. 
The goal of the experiment is simple: by increasing the dimension $d$ of the state in Eq.~\eqref{eq:etstate} through fine-graining, we want to certify entanglement of noisy quantum states, which is otherwise concealed by noise.\\\indent
To this end, we collect statistics about the state in two bases. The first measurement is in the same basis as the state in Eq.~\eqref{eq:etstate}. Projecting onto the time-bin states $\ket{i,j}$, with $i,j\in \{1,...,d\}$, is accomplished by recording the time of arrival of single photons with a detector and a precise clock, which constitutes a mulit-outcome measurement. The second measurement is more difficult as it must be performed in a superposition basis of the time-bin states.
This can be achieved by delaying the state $\ket{i}$ for a duration corresponding to $f$ time-bins and subsequently interfering it with the state $\ket{i+f}$. We realize this in our experiment by utilising a Franson interferometer \cite{Franson1989}, which employs an unbalanced interferometer for Alice and Bob respectively (see Fig.~\hyperref[fig:exp_setup]{\ref*{fig:exp_setup}(a)}). The long interferometer arm delays the state $\ket{i}$ relative to the state $\ket{i+f}$, which occupies the spatial path of the short interferometer arm. The second basis therefore projects onto the states $\tfrac{1}{\sqrt{2}}\left(\ket{i,j} + e^{i\phi}\ket{i+f,j+f}\right)$, where the phase $\phi$ is set by the sum of the two individual interferometer phases and $i,j\in \{1,...,d\}$. However, without active switching, this interferometer will 
also project onto the states $\ket{i,i+f}$ and $\ket{i+f,i}$, which are not interfering and thus must be discarded in coincidence post-selection. Since we investigate high-dimensional states, these non-interfering events are part of our state space and we may not simply discard them. We tackle this problem by employing a postselection-free Franson interferometer \cite{Strekalov1996}. In this scheme, polarization-entanglement is exploited to deterministically route the photon pairs in the Franson interferometer. This requires a hyperentangled source state \cite{Kwiat1997, Barreiro2005} in polarization and energy-time. We generate the additional polarization entanglement by bidirectionally pumping a nonlinear
crystal centered in a polarization Sagnac interferometer \cite{Kim2006,Fedrizzi2007}, which enables us to use the polarization DOF to switch between the two measurement basis in the time domain (see Appendix \hyperref[sec:appA]{A} for details). 
Our entanglement certification is based on a dimension-dependent entanglement witness $W(\rho_\text{ET},d)$, where, from Eq.~\eqref{eq:etstate} $\rho_\text{ET} = \mathrm{Tr}_\text{pol}\ket{\Psi}\bra{\Psi}$. The relation from the count statistics of the two measurements to the state $\rho_{ET}$ is rather involved and can be 
found in the Supplemental Material, Sec.~1 \cite{supplmat}. Here, it suffices to say that our state is entangled if $W(\rho_{ET}, d)>0$. 

\begin{figure*}
\centering
\includegraphics[width=1\linewidth]{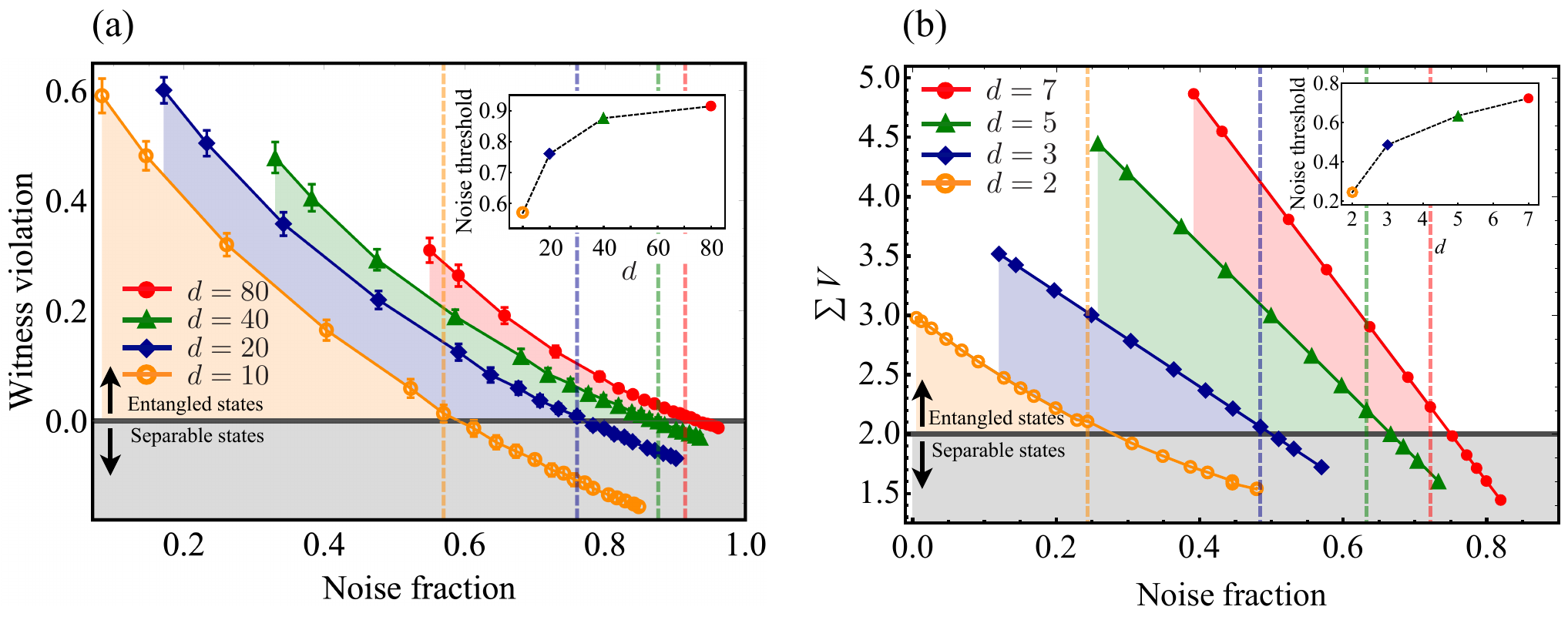}
\caption{Main results of our experimental demonstration of noise resistance for (a) energy-time entanglement and (b) OAM entanglement. Each plot depicts the violation strength of a suitably chosen entanglement witness against the noise fraction, i.e. the fraction of coincidence detections attributable to noise. In plot (a) the principal competition in achieving noise resistance is clearly visible. As the dimensionality is increased through fine-graining (Pathway I), more noise is induced (and thus the curves move to the right), while a higher noise resilience is achieved (thus the noise threshold also moves to the right). Plot (b) is qualitatively different, as it explores Pathway II. Instead of fine-graining,  more modes are included in the analysis which allow for an increased number of mutual unbiased bases to be measured and thus also show a higher noise threshold with increasing dimension. The error bars correspond to 3 standard deviations of the mean, calculated by propagating the Poissonian error in the photon-counting rates via a Monte Carlo simulation, see Supplemental Material, Sec.~4 \cite{supplmat}. In (b), the error bars are smaller than the data points.}
\label{fig:results}
\end{figure*}

We introduce increasing levels of external noise corresponding to a noise fraction $N\!F$ ranging from 0 to near-unity, in order to transition from a close-to-pure to a mixed state. Following Pathway I, we now fine-grain our state space to higher dimensions. The frame duration $F$ is fixed at 320 clock cycles and we discretise the frame in four ways according to
$F/d = t_d$ for $d\in\{10, 20, 40, 80\}$. This choice of dimensions depend on the imbalance of the Franson interferometer, and is detailed in Appendix \hyperref[sec:appA]{A}. 
Figure \hyperref[fig:results]{\ref*{fig:results}(a)} illustrates the scaling of the entanglement witness $W$ for different dimensions as the noise fraction $N\!F$ is increased. This increase is accomplished by incrementing the amount of external optical noise, whith the sequence of data points in each dimension corresponding to the same external noise levels. The noise threshold, which is the maximal $N\!F$ for which entanglement can be certified, increases with higher dimensions, indicating noise resilience (see inset). For $d=10,20,40,80$ the noise thresholds steadily increase from 0.57, 0.76, 0.86 to 0.93, respectively. As a consequence of fine-graining, the crosstalk between time-bins increases due to fundamental and technical limitations. This excess noise becomes relevant once the time-bin size is smaller than the timing resolution of the detectors, as is the case with $d=40$ and $d=80$. For these discretizations, the $N\!F$ is significantly increased even in the absence of external noise, indicated by the first data points in each dimension. Fine-graining at low external noise levels also reduces the witness violation, while for noise levels close to the noise threshold, fine-graining results in the recurrence of otherwise obscured entanglement. 

\newpage
\section{Orbital angular momentum  entanglement \\ (Pathway II)}

The second pathway to noise resilience takes advantage of the larger number of mutually unbiased bases in higher dimensions. Here, we explore this pathway using measurements of orbital angular momentum MUBs, for which precise measurements techniques have only recently been developed \cite{Bouchard2018measuring}. Mutually unbiased bases are an invaluable tool in many quantum information tasks, such as quantum state tomography, quantum cryptography, and entanglement certification. They consist of a set of orthonormal bases $\left\{\cal B_\alpha \right\}$, where ${\cal B_\alpha}=\left\{ |\psi_m^{(\alpha)} \rangle \right\}$, $m \in \{0,1,...,d-1 \}$ and $\alpha \in \{0,1,...,d\}$. Such a set is called mutually unbiased if and only if,

\begin{eqnarray}
\left| \langle \psi_m^{(\alpha)} | \psi_n^{(\beta)} \rangle \right|^2 = \delta_{\alpha \beta} \delta_{mn} + (1-\delta_{\alpha \beta})/d,
\end{eqnarray}
where $\delta_{i,j}$ is the Kronecker delta. In dimensions that are powers of prime numbers, it is known that there exists exactly $(d+1)$ MUBs. Surprisingly, for dimensions that are not powers of prime numbers, finding the number of MUBs and their elements remains an open problem~\cite{durt2010mutually}. For the case of prime dimensions and $\alpha \geq 1$, a MUB element is explicitly given by $| \psi_m^{(\alpha)} \rangle=\left(1/\sqrt{d}\right) \sum_{j=0}^{d-1} \left( \omega_d^{m} \right)^{d-j} \left(\omega_d^{-(\alpha-1)} \right)^{s_j} |j\rangle$, where $\omega_d = \exp(2 \pi i /d)$ and $s_j=j+...+(d-1)$. In the current experiment, we use the intensity flattening technique~\cite{Bouchard2018measuring} to measure the correlations of the photon pairs in all MUBs (see Appendix \hyperref[sec:appB]{B} for further details). The joint probability of Alice and Bob measuring states $|\psi_m^{(\alpha)} \rangle$ and $|\psi_n^{(\beta)} \rangle$ respectively, is given by $P^{(\alpha,\beta)}(m,n)$. For a complete set of joint measurements by Alice and Bob in bases ${\cal B}_\alpha$ and ${\cal B}_\beta$ respectively, we define the correlation visibility as $V^{(\alpha,\beta)}=\sum_{i=0}^{d-1} P^{(\alpha,\beta)}(i,i)$. Following the analysis of~\cite{spengler2012}, we obtain an upper bound for separable states by considering the sum of the visibilities over $k$ MUBs, i.e. $\sum_j^{k-1} V^{(j,j)} \leq 1+\frac{k-1}{d}$. In particular, for measurements in all $k=(d+1)$ MUBs, entanglement certification is achieved for $\sum_j^{k-1} V^{(j,j)} > 2$. Hence, in contrast to the case of energy-time entanglement described before, where detections are limited to measurements in two-dimensional subspaces but dimensions of up to 80, we are now able to fully characterize the generated states by performing high-dimensional projective measurements but we are limited to lower overall dimensions. However, this might be largely increased by using custom-tailored phase-matching \cite{svozilik2012high} or by considering the complete space of transverse spatial modes, namely radial modes along with azimuthal modes. 

As a starting point, we consider bi-dimensionally entangled OAM states of the form $(|1,-1\rangle + |-1,1\rangle)/\sqrt{2}$. 
Entanglement is certified by measuring correlations in all three MUBs in the two-dimensional space of OAM $|\ell =\pm 1 \rangle$. Environmental noise is steadily added by gradually increasing the intensity of the ambient light present in the lab, corresponding to a noise fraction $N\!F$ ranging from 0 to 0.8. Figure \hyperref[fig:results]{\ref*{fig:results}(b)} shows how the sum of visibilities ($\sum V$) in $d+1$ MUBs varies as a function of increasing noise fraction. Entanglement is always certified if $\sum V>2$, irrespective of dimension. For $d=2$, entanglement is certified for noise fractions up to 0.24. However, with increasing dimension, we are able to tolerate a higher noise fraction threshold, beyond which no entanglement can be certified (see inset). For $d=3$, 5, and 7, the noise fraction thresholds are 0.48, 0.63, and 0.72 respectively. The inset also shows that the noise threshold seems to be saturating as the dimension is increased. This is primarily due to the reduced fidelity of measurements in high dimensions, as well as our state moving further away from an ideal maximally entangled state as the dimension is increased. However, it is clear from our results that by increasing the state dimension, which in turn enables measurements in more bases, one can increase the resilience of entanglement to background noise. It is interesting to note that this could motivate the search for high-dimensional MUBs for any dimension, as communication systems should ideally be able to optimally operate beyond prime-dimensions.

\section{Discussion}
Our experimental results showcase the challenges and potential of overcoming noise through high-dimensional entanglement in quantum communication. 
While the necessary spatio-temporal entanglement is routinely generated in down-conversion, the real challenge is to encode information in these high-dimensionally entangled states. In other words, high-dimensional entanglement is already present in the workhorses of quantum communication, but routinely lost through coarse-graining and ignorance of modes. While this can be beneficial in removing noise from the signals, we observe a competition between two key factors: High-dimensional encoding increases the noise resistance as the dimension grows through the two pathways we identified, but also adds additional noise with increasing dimension. This is a competition that will ultimately always be won by noise, otherwise single photons could carry an infinite amount of information. The ultimate goal is finding the sweet spot, where the increased noise resistance still trumps the additional noise and thus realises a practical improvement in noisy entanglement distribution. What we show in our two experiments is that this sweet spot is actually beyond dimension two and thus defies conventional wisdom in the field, calling for the development of high-dimensional protocols across photonic platforms.
While we have used two different experiments to illustrate the two pathways to noise resistance separately, both pathways could in principle be realised simultaneously in the same experiment. If one had access to multiple MUB measurements in time-bins or multi-outcome measurements in the spatial domain, one could harness both pathways, leading to an increase in measurable dimensionality and as a consequence, higher noise thresholds.

Our method of adding external noise, namely by fixing a constant luminosity light source close to our detectors, is a fairly realistic model of noise that captures the decreased signal-to-noise ratio in long-distance quantum communication, where detector dark counts start dominating the distance-attenuated single-photon pairs. 
On the other hand, our experiments also simulate daylight conditions for free-space quantum communication \cite{Peloso2009, Liao2017}, where background photons will trigger accidentals in the very same way as our artificial lamps do. In both of these scenarios, the most detrimental noise in the quantum channel is white, which motivated us to employ noise sources of this characteristic in our experiments.

The most remarkable outcome of this study is the fact that we demonstrate the possibility to certify entanglement that was otherwise obscured. In other words, entanglement really was able to overcome physical noise in the implementation and reveal itself by going to higher dimensional encodings. We would like to note that this is not only a proof-of-principle implementation, but it is ready to be also directly adopted for long distance or free-space quantum communication \cite{Steinlechner:2017,Krenn:2016fk}. At least for the energy-time experiment we could use the exact same setup, whereas for the OAM experiment we would require a multi-outcome measurement, such as the recently developed spatial mode sorter \cite{Mirhosseini:2013em}. With the current single-outcome measurements, every element/dimension we add will experience the same environmental noise [since it directly couples to the single-mode fiber (SMF)], thus unfavourably influencing the competition between noise and entanglement, with the total noise fraction increasing at the same rate as the additional noise robustness. The noise fraction we measured nonetheless proves that, if one had a measurement technique where the noise distributes over multiple channels, we would have a tremendously increased resistance to physical noise outside of laboratory settings.

The obvious next challenge is the development of quantum communication protocols that make direct use of high-dimensional encodings. The fact that entanglement can be certified under extremely noisy background conditions motivates the question of whether such noisy entanglement can indeed be used to certify security of QKD or aid in other quantum information tasks. It has recently been proven that every entangled state, no matter how noisy, provides an advantage in entanglement-assisted classical communication \cite{StevenTreeler}. In addition, every noisy entangled state also provides an advantage for the task of channel discrimination \cite{MarcusPianus}. We hope that this study spurs further investigation into information theoretic protocols based on high-dimensional and noisy entangled states, which can be distributed in regimes where no qubit communication is possible.

\section*{Acknowledgements}
We thank Jessica Bavaresco for helpful discussions on OAM entanglement certification and data analysis.
F.Brandt, M.H., M.M., R.F. and Y.G. acknowledge funding from the Austrian Science Fund (FWF): Y879-N27, I3053-N27, P31339-N27.
M.M.~acknowledges support from the QuantERA ERA-NET co-fund (Austrian Science Fund (FWF): I3773-N36) and from the UK Engineering and Physical Sciences Research Council (EPSRC) (EP/P024114/1). F.Bouchard acknowledges the support of the Vanier Canada Graduate Scholarships Program and the Natural Sciences and Engineering Research Council of Canada (NSERC). R.F.~acknowledges the support of the Academy of Finland (Competitive Funding to Strengthen University Research Profiles - decision 301820 and Photonics Research and Innovation Flagship - decision 320165). We acknowledge funding from the Austrian  Research  Promotion  Agency (FFG) Quantenforschung und -technologie (QFTE) Contract 870003, Austrian Science and Applications Programme (ASAP) Contract 854022 and Contract 866025 and ESA European Space Agency Contract 4000112591/14/NL/US.

S.E. and F.Bouchard contributed equally to this work; M.M., R.U.~and M.H.~conceived the project; S.E., L.B.~and O.K.~designed and developed the energy-time entanglement experiment under the guidance of F.S.~and R.U.; F.Bouchard~and F.Brandt~designed and developed the orbital-angular-momentum entanglement experiment under the guidance of R.F., M.M.~and M.H.; M.H.~and Y.G.~established the entanglement certification methods; S.E., F.Bouchard, R.F., M.M, Y.G.~and M.H.~wrote the first draft of the manuscript; All authors discussed the results and reviewed the manuscript; M.H.~supervised the whole project.

\appendix
\section*{APPENDIX A: Energy-time entanglement experiment}
\label{sec:appA}
The experimental setup can be divided into a hyperentangled photon pair source, a Franson interferometer consisting of two imbalanced polarizing Mach-Zehnder interferometers (PMZI) and a detection- and time tagging-unit. Our source is based on SPDC in a 20 mm-long periodically poled potassium titanyl phosphate (ppKTP) crystal designed for type-II quasi-phase-matching. A grating-stabilized photodiode (Toptica DL pro) emitting at a wavelength of $\unit[405]{nm}$ is generating the pump field for the SPDC. Due to a narrow pump bandwidth of $\Delta \nu_{\text{FWHM}}\sim\unit[500]{kHz}$, the down-converted signal and idler fields are energy-time-entangled within a coherence time of $t_{\text{coh}} =1/(\pi \Delta \nu_{\text{FWHM}}) \sim\unit[636]{ns}$. The ppKTP crystal is temperature-tuned to produce wavelength-degenerate photon pairs at $\unit[810]{nm}$. In order to obtain polarization entanglement, the crystal is bidirectionally pumped in the center of a polarization Sagnac interferometer \cite{Kim2006,Fedrizzi2007}. After $\unit[3]{-nm}$ bandpass filtering and single-mode coupling,  we detect an entangled photon pair rate of $\unit[15]{kcps}$ per mW of pump power with a heralding efficiency of $\unit[20]{\%}$ in both signal and idler modes.

The single photons are then guided to two bulk optics PMZIs  
with an imbalance between long and short interferometer arm of $\unit[2.67]{ns}$. The imbalance of the two PMZIs is matched up to the correlation length of the photon pairs ($\sim \unit[800]{\mu m}$). By adjusting the phases $\phi_\text{A/B}$ of Alice's/Bob's PMZI, we see Franson interference with a phase of $\phi_\text{Franson}=\phi_\text{A} + \phi_\text{B}$. All of our measurements in the superposition or Franson basis are performed at maximal Franson interference contrast ($\phi_\text{Franson} = 0$ or $ \pi$), which requires phase-stability of the PMZIs over the measurement time.
Active phase stabilisation of both PMZIs is achieved by a control loop of a Piezo actuator displacing an interferometer mirror and the difference signal from two photodiodes indicating the interference contrast. This interference signal is provided by a $\unit[780.241]{-nm}$ stabilisation laser (Toptica DL Pro) propagating in the same spatial interferometer mode as the single photons.  It is injected into the PMZIs via the unused port of the first polarizing beam splitter (PBS) and measured at the output of the unused port of the second PBS, where the polarisation contrast is measured by fast photodiodes (Thorlabs - DET 10 A/M) in a polarisation basis conjugate to the polarisation basis defined by the PMZIs. The stabilisation laser is frequency-locked to a hyperfine transition of  ${}^{85}\text{Rb}$, obtained by saturated absorption spectroscopy, resulting in a wavelength stability of $\sim\unit[0.6]{fm/min}$.

We choose the measurement basis in the energy-time domain by changing the polarization measurement basis after the PMZI, effectively switching the interferometer on or off by erasing or revealing the interferometer path information. Performing a polarization measurement in the PMZI-defined rectilinear basis corresponds to a measurement in the computational basis, while projecting the photons in a mutually unbiased polarization basis corresponds to a measurement in the Franson basis (see Supplemental Material, Sec.~3 \cite{supplmat} for a stringent formal treatment). Noise is optically added to the measurement data by means of fine-adjustable light emitting diodes (LEDs) powered by a battery, ensuring time-invariant noise generation. 
We detect both polarization components on Alice's (detectors A0 and A1) and Bob's (detectors B0 and B1) side by means of multimode-coupled single-photon avalanche diodes (Excelitas SPCM-800-11) with a measured FWHM timing jitter of $<\unit[800]{ps}$ between two detectors. The detection events are time-tagged employing a time to amplitude converter (AIT TTM8000) with a clock resolution of \unit[82.3]{ps}.

Post-processing of the time-tagged data is realized by binning the detection events of each channel into dimension-dependent time-bins of duration $t_d = F/d$, where $F$ is the duration of one frame.   
Since the imbalance of our interferometers is fixed and corresponds to 32 clock cycles, 
only time-bin durations which obey $f\cdot t_d = 32$ clock cycles
give rise to well-defined Franson interference $\ket{i,i} + e^{i\phi_\text{Franson}}\ket{i+f,i+f}$, where $f$ is an integer corresponding to the time-bin shift. To this end, in order to see interference, we investigate dimensions which satisfy $d = \frac{f\cdot F}{32}$ for integer $f$ and $d$. For our setup parameters and for a time-frame duration of $F=320$ clock cylces this corresponds to  $d\in \{10, 20, 40, 80\}$. 

The discretizations to different dimensions are performed on the same set of measurement data. Since we are tracking photons emitted from a photon pair source, our state space is intrinsically bipartite, and 
only those time-frames which contain exactly one detection event on Alice's side and exactly one on Bob's side are kept; all others are discarded (e.g. no detection event in Alice's and 1 detection event in Bob's detectors per frame). The detection events which are kept are then sorted into count matrices pertaining to the detectors that clicked
(A0-B0, A1-B1, A0-B1, A1-B0).
These 4  matrices in both measurement bases are used to reconstruct the part of the state $\rho_\text{ET}$ required in the subsequent entanglement certification. \\
Since the timing-jitter of the detectors is one order of magnitude greater than the clock resolution of the timetagger, our overall timing resolution is dominated by the detector jitter. Therefore, crosstalk errors between time-bins will sharply increase once the time-bin duration  $t_d$ is on the order of the timing-jitter of the detector, which is the case for $t_d = \unit[8]{\text{clock cycles}} = \unit[658.4]{ns}$.

The witness used to certify entanglement was derived using the entropy vector formalism in \cite{HuverVincente2013}. For each dimension $d$, the underlying state $\rho_\text{ET}$ is not separable (i.e. entanglement is certified) if $W(\rho_\text{ET}, d)>0$, where

\begin{align} \tag{A1}
\begin{split}
& W(\rho_\text{ET}, d) :=
	\sum_i^{d-f} |\bra{ii} \rho_\text{ET}\ket{i+f , i+f}|\\
	&- \sqrt{\bra{i , i+f} \rho_\text{ET}\ket{i , i+f}\bra{i+f , i} \rho_\text{ET}\ket{ i+f , i}}.
	\end{split}
	\end{align}
	
In order to compute the witness, one must reconstruct the underlying density matrix elements of $\rho_\text{ET}$ from the experimental count matrices. These also depend on the polarisation degree of freedom, due to the use of a postselection-free Franson interferometer. Details of how to compute the witness from the count matrices produced by the experiment can be found in the Supplemental Material, Sec.~1 \cite{supplmat}.\\

\section*{Appendix B: OAM entanglement - experiment}
\label{sec:appB}
We generate pairs of photons entangled in the orbital-angular-momentum (OAM) degree of freedom by pumping a 5~mm long ppKTP crystal quasi-phase matched for type II SPDC. We use a 405~nm diode laser (Toptica iBeam Smart 405 HP) that is coupled to a single-mode optical fiber to ensure the best possible transverse coherence and mode profile, which is essential to obtain high-dimensionally entangled pairs of photons. The UV beam is focused by a 500~mm lens to a spot size of 430~$\mu$m (1/$e^2$ beam diamater) at the ppKTP crytal. We similarly temperature tune this crystal to produce pairs of wavelength-degenerate, orthogonally polarized photons at 810~nm. The photon pairs are recollimated by a 300~mm lens. This time the polarization DOF of the photons is solely used to deterministically split the photons at a polarization beam splitter, such that their spatial mode can be measured independently of each other. The photons are then made incident on phase-only spatial light modulators (Holoeye PLUTO), where a combination of computer-generated holograms and single-mode fibers (SMFs) are used to perform a generalized projective measurement in the OAM state space. Finally, the photons are detected by avalanche photodiodes and coincidence measurements are recorded within a coincidence time window of 5~ns using a custom-built logic. In the computational basis, measurements of photonic OAM may be accomplished by displaying a hologram generating the opposite OAM value, thus resulting in an outgoing beam with a flat wavefront with an OAM value of $\ell=0$ that will couple efficiently to the SMF using a 10-X microscope objective. This technique is also known as phase-flattening and has been widely used to measure the OAM content of an unknown beam~\cite{mair2001entanglement}. However, in order to certify entanglement, it is necessary to perform measurements in additional bases besides the computational (OAM) basis, which leads to more complex mode structures (see Supplemental Material, Sec.~5 \cite{supplmat}). Thus, a more elaborate measurement scheme is required to accurately measure the general OAM state of the experimentally generated entangled pairs. We use a recently introduced technique called intensity flattening \cite{Bouchard2018measuring}, that allows one to measure any arbitrary transverse spatial mode of light, including modes in any mutually unbiased basis of OAM. Although lossy, this technique yields extremely high detection fidelities. Using this source, after taking into account the lossy intensity masking holograms implemented at the spatial light modulators, we achieve coincidence count rates of 500 Hz in the fundamental Gaussian mode, 1000 Hz in the first-order OAM modes ($\ell = \pm$ 1), 700 Hz in the second-order OAM modes ($\ell = \pm$ 2), and 400 Hz in the third-order OAM modes ($\ell = \pm$ 3). The associated singles count rates are given by 13 kHz,  20 kHz, 15 kHz, and 11 kHz.

\nocite{HuverVincente2013,HuberLobetVincente2013}

\providecommand{\noopsort}[1]{}

\clearpage
\onecolumngrid

\section*{Supplementary Information}

\subsection{Pathway I - Entanglement witness derivation}
\label{sec:ent_witness}
In order to certify entanglement in the energy-time experiment (Pathway I), we appealed to the entropy vector method, first introduced in \cite{HuverVincente2013} and elucidated in \cite{HuberLobetVincente2013}, which was used to investigate the structure of multipartite entanglement. 
Consider the pure state $\rho$ consisting of $n$ parties. Let $r$ denote a particular subset of parties, i.e. a subset of
$\{1, 2,\cdots , n\}$, and let $\bar{r}$ denote the complement. The set $\mathcal{R}$ denotes a further subset of $r$. 
For pure states $\rho$, the components of the linear entropy vector $\vec{S}_L$ are given by  
\begin{align}\label{eq:vecbound}
S_L^j (\rho )\ge -\log_2\bigg(
1 - \frac{W_j(\rho, C, \mathcal{R})^2}{2}
\bigg)
\end{align}
where the witness for the $j-$th component of the vector is given by
\begin{align}\label{eq:witness}
W_j(\rho, C, \mathcal{R}): = \frac{1}{\lvert \sqrt{C} \rvert} \sum_{\eta, \eta^\prime \in  C}
\bigg(
\bra{\eta} \rho\ket{\eta^\prime}- \min_{\{r_m\}\in \mathcal{R}}
\sum_{m=1}^j
 \sqrt{\bra{\eta_{r_m}} \rho\ket{\eta_{r_m}}\bra{\eta'_{r_m}} \rho\ket{\eta'_{r_m}}
 }
\bigg)
\,.
\end{align}
Here $\eta$ is a multi-index (e.g. the triple $110$ for a tripartite qubit state), and the pair $(\eta_{r_m}, \eta'_{r_m})$ is the pair $(\eta_{}, \eta'_{})$ with the indicies of the  $r_m$ subset of parties exchanged. $C$ is a set of indices over which the witness runs, and can be chosen as desired. Intuitively, this witness sums particular off-diagonal terms from the density matrix, and penalises those on the diagonal. As was shown in \cite{HuverVincente2013, HuberLobetVincente2013}, if all entries of $\vec{S}_L $ are non-zero then the $n$-partite state cannot be written as a convex combination of separable states
\begin{align}
\rho\neq \sum_i p_i \rho_{r_i}\otimes \rho_{\bar{r}_i}
\end{align}
i.e. the state has no separable decomposition, which implies that it is entangled. 
Since we are working to verify the entanglement of a bipartite state $\rho_{AB}$, this trivially selects the set $\mathcal{R} = \{A\}$\footnote{Or equivalently $\{B\}$ under symmetry.} such that $j= 1$ in Eq.~\eqref{eq:witness} and the sum and minimisation vanish. Thus,  $\vec{S}_L $ has one component, and if it is non-zero then
\begin{align}
\rho_{AB}\neq \sum_i p_i \rho_{A_i}\otimes \rho_{B_i}
\,
\end{align}
and entanglement is certified. 
 From Eq.~\eqref{eq:vecbound} it can be seen that each witness $W_j$ provides a lower bound on each component $S^j_L$ of the entropy vector; thus a necessary condition for $\rho_{AB}$ to be entangled is that $W_1(\rho_{AB}, C, \{A\}) > 0$. \\\\
 In principle, in order to get the best witness out of Eq.~\eqref{eq:witness} one may play with the set $C$ in order to maximise the expression. In this work we do not perform such an optimisation and simply take the set to be $C =\{ ((i,i) \,,\, (i+1, i+1))\}_1^{d-1}$, since we anticipate the produced state to be close to $\ket{\Phi^+}\bra{\Phi^+} =\sum_{ij}\frac{1}{d}\ket{ii}\bra{jj}$. This move means that we present a lower bound on the witness in Eq.~\eqref{eq:actualwitness} which, as we shall show, is sufficient to verify entanglement anyway. With these definitions in place the witness becomes 
\begin{align}\label{eq:actualwitness}
W_1(\rho) := \frac{1}{\sqrt{d-1-f}}
\bigg(
	\sum_i^{d-f} \bra{ii} \rho\ket{i+1 , i+1}- \sqrt{\bra{i , i+1} \rho\ket{i , i+1}\bra{i+1 , i} \rho\ket{ i+1 , i}}
	\bigg)
	\,,
\end{align}
where we have dropped the subscript $AB$ for convenience. 
From this expression, it is clear that the maximum value of the witness is achieved on the maximally entangled state $ W_1(\ket{\Phi^+}\bra{\Phi^+})  = \frac{d-1}{d\sqrt{d-1}}$.
Moreover, it is clear that if we take $C =\{ ((i,i) \,,\, (i+f, i+f))\}_1^{d-1}$ for integer $f$, then
\begin{align}\label{eq:shiftedwitness}
W_1(\rho) := \frac{1}{\sqrt{d-1}}
\bigg(
	\sum_i^{d-1} \bra{ii} \rho\ket{i+f , i+f}- \sqrt{\bra{i , i+f} \rho\ket{i , i+f}\bra{i+f , i} \rho\ket{ i+f , i}}
	\bigg)
	\,,
\end{align}
is also a valid entanglement witness. Due to the particulars of the experimental setup, we will work with the above expression. In order to compute the value of the witness on the state  that is produced by experiment, we must reconstruct the density matrix elements of $\rho$ from the count matrices. In short, the first term in Eq.~\eqref{eq:shiftedwitness} is not a number we have direct access to, and thus it must be computed (in fact bounded) from the data; on the other hand, the term in the square root is measured and can be extracted directly from the experimental data. \\

The time-bin entanglement experiment provides count matrices for detection clicks produced by making one of two measurements on Alice and Bob's photons. These are either the `computational basis', or the `Franson basis', alluded to in the main text, which, for convenience, we refer to from here on as the `horizontal' (HV) and `diagonal' (DA) bases. The same local measurement is always performed on each side, i.e., either  $M^{\text{DA}}_{A}\otimes M^{\text{DA}}_{B}$ or $M^{\text{HV}}_{A}\otimes M^{\text{HV}}_{B}$, is measured. The polarisation degree of freedom of the photon pairs in the post-selection free Franson interferometer is used as a proxy for detecting the time-bin that Alice and Bob's photons landed in. Thus, a measurement in either of these bases corresponds to one of $4$ possible events, namely a click in one of the detector pairs $\{A0B0, A0B1, A1B0, A1B1 \}$. 
The task is to compute the entanglement witness on the energy-time entangled state $W_1(\rho_{ET})$. For polarisation measurements in the HV basis, the full  polarisation and time-bin entangled state $\rho_{P,ET}$ is related to the experimental count matrices in the following way:
\begin{align}
\begin{split}
    \langle HH|\langle ij | \rho_{P,ET}|ij\rangle|HH\rangle&=\frac{\langle i|HV_{A0B0}|j\rangle}{N_1}\\
     \langle HV|\langle ij | \rho_{P,ET}|ij\rangle|HV\rangle&=\frac{\langle i|HV_{A0B1}|j+f\rangle}{N_1}\\
      \langle VH|\langle ij | \rho_{P,ET}|ij\rangle|VH\rangle&=\frac{\langle i+f|HV_{A1B0}|j\rangle}{N_1}\\
       \langle VV|\langle ij | \rho_{P,ET}|ij\rangle|VV\rangle&=\frac{\langle i+f|HV_{A1B1}|j+f\rangle}{N_1}
       \end{split}
\end{align}
where  $N_1=\sum_{k,l=0}^1\sum_{i=1-k}^{d-1-k}\sum_{j=1-l}^{d-1-l}\langle i|HV_{AkBl}|j\rangle$ is the normalisation over all counts and $f$ is a known delay due to the interferometer imbalance \footnote{\begin{minipage}[t]{\textwidth}Note that $f = f(d)$ depends on the dimension that the frame $F$ is divided into, and the experimental parameters are always \\ chosen such that $f$ is an integer.\end{minipage}}. From this, we are able to determine the diagonal elements of the energy-time matrix $\rho_{ET}$ by eliminating (i.e. summing over) the polarisation,
\begin{align}\label{eq:HVdef}
          \langle ij | \rho_{ET}|ij\rangle
            =\frac{\sum_{k,l=0}^1\langle i+f, k|HV_{AkBl}|j+f,l\rangle}{N_1}
            \,.
\end{align}

To reconstruct the density matrices from the count matrices pertaining to measurements in the DA basis on has to do a little more work. The DA basis performs measurements   due to the non-orthogonality of projections onto neighbouring bin pairs. We proceed by showing the relation for the first two count matrices,  

\begin{align}
\tfrac{1}{4}\Big(\bra{Hi} +\bra{V i+f}\Big)_A
    \Big(\bra{Hj}+ \bra{V j+f} \Big)_B \rho_{P,ET}
        \Big(\ket{Hi} +\ket{V i+f}\Big)_A   
            \Big(\ket{Hj}+ \ket{V j+f}\Big)_B 
            & = \frac{\langle i|DA_{A0B0}|j\rangle}{N_2} \\
    \tfrac{1}{4}\Big(\bra{Hi} -\bra{V i+f}\Big)_A
\Big(\bra{Hj}- \bra{V j+f} \Big)_B \rho_{P,ET}
\Big(\ket{Hi} -\ket{V i+f}\Big)_A   
    \Big(\ket{Hj}- \ket{V j+f}\Big)_B 
    & = \frac{\langle i|DA_{A1B1}|j\rangle}{N_2}
\end{align}

Let $p_+=\langle HH|\rho_P|HH\rangle+\langle VV|\rho_P|VV\rangle$ and $p_-=\langle HV|\rho_P|HV\rangle+\langle VH|\rho_P|VH\rangle$. We take $i = j$, and assume that $\rho_{P,ET}=\rho_P\otimes\rho_{ET}$. This being justified by the fact that across the partition between the $2$-dimensional polarization and $d$-dimensional time-bin states on Alice and Bob, the maximum Schmidt number is $2$. Moreover, since we expect the time-bin state to be close to maximally entangled, by monogamy we should expect close to zero entanglement across the partition with polarization. With this in place, we have that the sum of these terms is   
\begin{align}
\begin{split}
   \sum_{i=0}^{d}\frac{\langle i|DA_{A0B0}|i\rangle}{N_2}+\frac{\langle i|DA_{A1B1}|i\rangle}{N_2}=\\
   { \sum_{i=0}^{d}\frac{1}{2}(p_+\langle ii|\rho_{ET}|ii\rangle+p_-\langle i,i+f|\rho_{ET}|ii+f\rangle +p_-\langle i+f,i|\rho_{ET}|i+f,i\rangle+p_+\langle i+f,i+f|\rho_{ET}|i+f,i+f\rangle)}\\
   +\Re e[\langle HH|\rho_P|VV\rangle\langle ii|\rho_{ET}|i+f,i+f\rangle]+\Re e[\langle HV|\rho_P|VH\rangle\langle i,i+f|\rho_{ET}|i+f,i\rangle]+c
   \end{split}
\end{align}
where $N_2=N_1\eta_{HWP}^2$ and $\eta_{HWP}$ is the efficiency of the half-wave-plate used for selecting the basis. Note that the constant $c$ enters due to boundary effects where terms at the beginning and end of the summation do not cancel due to the shift $f$.
The other two count matrices are calculated to be
\begin{align}
\tfrac{1}{4}\Big(\bra{Hi} +\bra{V i+f}\Big)_A
    \Big(\bra{Hj}- \bra{V j+f} \Big)_B \rho_{P,ET}
        \Big(\ket{Hi} +\ket{V i+f}\Big)_A   
            \Big(\ket{Hj}- \ket{V j+f}\Big)_B 
            & = \frac{\langle i|DA_{A0B1}|j\rangle}{N_2} \\
    \tfrac{1}{4}\Big(\bra{Hi} -\bra{V i+f}\Big)_A
\Big(\bra{Hj}+ \bra{V j+f} \Big)_B \rho_{P,ET}
\Big(\ket{Hi} -\ket{V i+f}\Big)_A   
    \Big(\ket{Hj}+ \ket{V j+f}\Big)_B 
    & = \frac{\langle i|DA_{A1B0}|j\rangle}{N_2}
    \,.
\end{align}

The sum of which is 
\begin{align}
   \sum_{i=0}^{d}\frac{\langle i|DA_{A0B1}|i\rangle}{N_2}+\frac{\langle i|DA_{A1B0}|i\rangle}{N_2}=\nonumber\\{ \sum_{i=0}^{d}\frac{1}{2}(p_+\langle ii|\rho_{ET}|ii\rangle+p_-\langle i,i+f|\rho_{ET}|i,i+f\rangle +p_-\langle i+f,i|\rho_{ET}|i+f,i\rangle+p_+\langle i+f,i+f|\rho_{ET}|i+f,i+f\rangle)}\nonumber\\-\Re e[\langle HH|\rho_P|VV\rangle\langle ii|\rho_{ET}|i+f,i+f\rangle]-\Re e[\langle HV|\rho_P|VH\rangle\langle i,i+f|\rho_{ET}|i+f,i\rangle]+c
   \,.
\end{align}
Taking the following, particular, linear combination of these matrices
\begin{align}
   \sum_{i=0}^{d-f}\frac{\langle i|DA_{A0B0}|i\rangle}{N_2}+\frac{\langle i|DA_{A1B1}|i\rangle}{N_2}-\frac{\langle i|DA_{A0B1}|i\rangle}{N_2}-\frac{\langle i|DA_{A1B0}|i\rangle}{N_2}=\nonumber\\
    \sum_{i=0}^{d-f}2\Re e[\langle HH|\rho_P|VV\rangle\langle ii|\rho_{ET}|i+f,i+f\rangle]+2\Re e[\langle HH|\rho_P|VV\rangle\langle i,i+f|\rho_{ET}|i+f,i\rangle]
\end{align}
 we are able to bound the first term appearing in the witness in Eq.~\eqref{eq:actualwitness}
\begin{align}
         \sum_{i=0}^{d-f} |\langle ii | \rho_{ET}|i+f,i+f\rangle|
         \geq
            \sum_{i=0}^{d-3-f}(\frac{\langle i|DA_{A0B0}|i\rangle}{N_2}+\frac{\langle i|DA_{A1B1}|i\rangle}{N_2}-\frac{\langle i|DA_{A0B1}|i\rangle}{N_2}-\frac{\langle i|DA_{A1B0}|i\rangle}{N_2}\nonumber\\
          -2\sqrt{\langle HV | \rho_{P}|HV\rangle\langle VH| \rho_{P}|VH\rangle}\sqrt{\langle i+f,i | \rho_{ET}|i+f,i\rangle\langle i,i+f | \rho_{ET}|i,i+f\rangle})
\end{align}
where we used $\langle HH|\rho_P|VV\rangle\leq\frac{1}{2}$, $|z|\geq\Re e[z]$, and $ |\langle ij | \rho|kl\rangle|\leq\sqrt{\langle ij | \rho|ij\rangle|\langle kl | \rho|kl\rangle|}$. Assuming the worst case algebraic bound $\sqrt{\langle HV | \rho_{P}|HV\rangle\langle VH| \rho_{P}|VH\rangle} \le \frac{1}{2}$, we have 
\begin{align}
    W_1(\rho_{ET}) \ge \frac{1}{\sqrt{d-1}}\sum_{i=0}^{d-3-f} 
    \bigg( \frac{\langle i|DA_{A0B0}|i\rangle}{N_2}+\frac{\langle i|DA_{A1B1}|i\rangle}{N_2}-\frac{\langle i|DA_{A0B1}|i\rangle}{N_2}-\frac{\langle i|DA_{A1B0}|i\rangle}{N_2}\nonumber\\
          -\sqrt{\langle i+f,i | \rho_{ET}|i+f,i\rangle\langle i,i+f | \rho_{ET}|i,i+f\rangle}
    \bigg)
\end{align}
Our final step is to substitue the terms under the square root with their definition from Eq.~\eqref{eq:HVdef} and argue that since we are looking for a positive violation $W_1(\rho_{ET}) > 0$ to witness entanglement, then the pre-factor is irrelevant. The final witness we test on the experimental data is thus
\begin{align}
    W_1(\rho_{ET}) \ge 
        &\sum_{i=0}^{d-f} 
        \frac{1}{N_2}\bigg(
 {\langle i|DA_{A0B0}|i\rangle}+{\langle i|DA_{A1B1}|i\rangle}-{\langle i|DA_{A0B1}|i\rangle}-{\langle i|DA_{A1B0}|i\rangle}
 \bigg)\nonumber\\
          &\qquad\qquad -2\sqrt{\frac{1}{N_1}
            \sum_{k,l=0}^1 \sum_{m,n=0}^1 \langle i+f,i | HV_{AlBk}|i+f,i\rangle\langle i+f,i+f |  HV_{AmBn}|i+f,i+f\rangle}
    \, ,
\end{align}
where $f$ is the adjustment we make to the witness in lieu of the inteferometer imbalance. In all, we test four dimensions (i.e. discretisations of the frame), which are summarised below.

\begin{table}[H]
\begin{center}
\caption{Different time-frame discretisations with corresponding f-shifts}

\begin{tabular}{c|c|ccl}
{\textbf{Dimension $d$}} & \textbf{\begin{tabular}[c]{@{}c@{}}Time-bin duration\\  in clock cycles (ns)\end{tabular}} & { \textbf{f-shift}} &                      &  \\ \cline{1-3}
$10$                     & $32 \,(2.62)$                              & $1$              &                      &  \\
$20$                     & $16 \,(1.31)$               & 2              &                      &  \\
$40$                     & $8 \,(0.66)$                  & 4              &                      &  \\
$80$                     & $4 \,(0.33)$                 & $8$              & \multicolumn{1}{l}{} & 
\end{tabular}
\label{tab:discr}
\end{center}
\end{table}

\subsection{Pathway I and II - Definiton of the Noise fraction $N\!F$}
In the main text we quantify the level of noise via the noise fraction $N\!F$, which is defined as the total noise divided by the total signal in our detectors. Here, we elaborate on this quantity and discuss how it is extracted from the measurement data of both experiments. Suppose Alice chooses to measure in the basis $\alpha$, while Bob chooses to measure in the basis $\beta$. The number of two-photon events with measurement outcome $m \in \{0,1,...,d-1\}$ for Alice and $n\in \{0,1,...,d-1\}$ for Bob in their respective bases is given by $N^{(\alpha,\beta)}(m,n)$. Let us first define the $N\!F$ used in Pathway I. For this Pathway, we quantify the noise exclusively in the computational, or time of arrival (TOA) basis as 
\begin{equation}
    N\!F=\frac{\displaystyle\sum_{m=0}^{d-1}\displaystyle\sum_{\substack{n=0 \\ n\neq m}}^{d-1}N^{(\text{TOA},\text{TOA})}(m,n)}{\displaystyle\sum_{m=0}^{d-1}\displaystyle\sum_{n=0}^{d-1}N^{(\text{TOA},\text{TOA})}(m,n)}.
\end{equation}
In the numerator, all uncorrelated or off-diagonal elements of $N^{(\text{TOA},\text{TOA})}(m,n)$ are summed, while the denominator is the sum over all two-photon events. Pathway II, on the other hand, harnesses all mutually unbiased bases (MUBs) in each dimension $d$, which means no basis is distinguished. Hence, we average over the noise fractions of all $d+1$ MUBs, yielding  
\begin{equation}
    N\!F=\frac{1}{d+1}\displaystyle\sum_{\alpha=0}^{d}\left(\frac{\displaystyle\sum_{m=0}^{d-1}\displaystyle\sum_{\substack{n=0 \\ n\neq m}}^{d-1}N^{(\alpha,\alpha)}(m,n)}{\displaystyle\sum_{m=0}^{d-1}\displaystyle\sum_{n=0}^{d-1}N^{(\alpha,\alpha)}(m,n)}\right).
\end{equation}

\newpage
\subsection{Pathway I - Discretization and correlation measurements}
\label{sec:bin_frame}
\begin{figure}[H]
\centering
\includegraphics[width=1\linewidth]{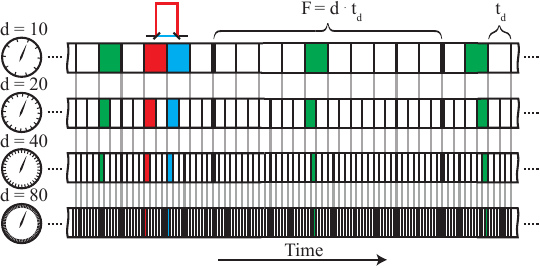} 
\caption{Discretization of the time domain in time-frames and time-bins of duration $t_d$ with examples of single-photon detection events (in green). The time-frame duration $F$ is kept constant for all dimensions $d$. Higher dimensions are therefore implemented by decreasing the time-bin duration according to $t_d=\frac{F}{d}$, which leads to Franson interference (in blue-red) between well-defined time-bins irrespective of the dimension.}
\label{fig:bin_frame}
\end{figure}

Pathway I is realized by fine-graining the time of arrival of single photons. To this end, the arrival time is discretized in time-bins of duration $t_d$, and the time-bin number of a photon detection is recorded. The time-bins are numbered from 1 to $d$, adding up to a time-frame of duration $F=d \cdot t_d$, which is kept constant at \unit[26.2]{ns} for our experiment. A fixed time-frame duration ensures that the noise rate per frame is constant irrespective of the dimension. There are two constraints on the time-bin duration: Firstly, $t_d$ must be a multiple of the clock cycle of our time to amplitude converter (\unit[82.3]{ps}), and secondly, the fixed imbalance of the Franson interferometer must be a multiple of $t_d$ (see Fig.~\ref{fig:bin_frame}). These constraints yield time-bin durations and f-shifts listed in Table \ref{tab:discr}.

\begin{figure}[H]
\centering
\includegraphics[width=1\textwidth]{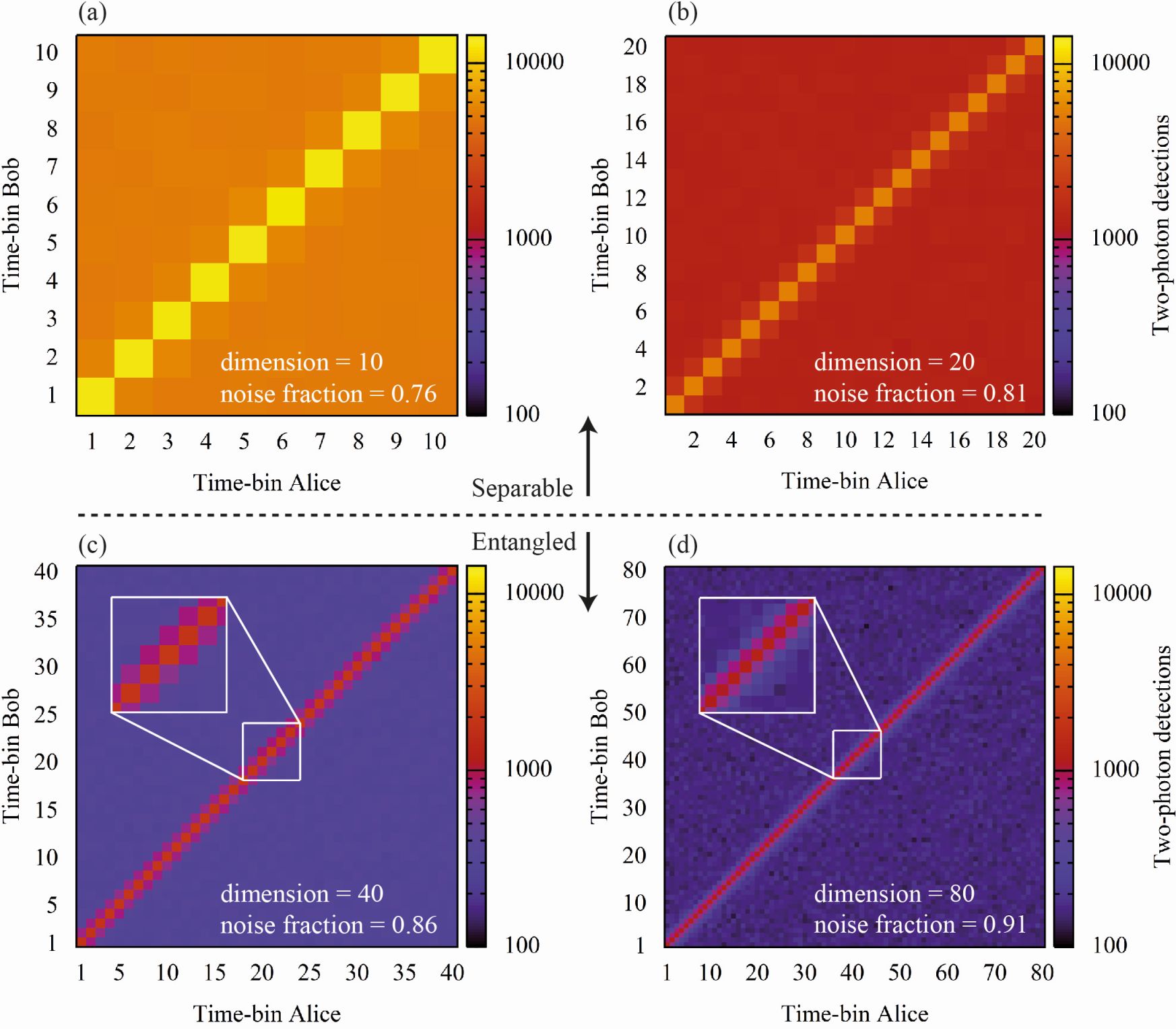} 
\caption{Two-photon detection events in the time-domain integrated over 4 minutes. The correlated events on the diagonal primarily arise from energy-time entangled photon pairs emitted from a SPDC source, while the off-diagonal elements can be attributed to high levels of external noise. Photon detection-events are discretized in time-bins of dimension $d =$  (a) 10, (b) 20, (c) 40 and (d) 80. 
The time-frame has a constant duration throughout all dimensions, which leads to a decrease in time-bin duration with increasing $d$. The plots are generated from detection events between detectors A0 and B1 from a measurement in the Franson basis and at a constant external noise level of $\sim \unit[400]{kcps}$ per detector. Increasing crosstalk due to timing-jitter leads to an increase in the noise fraction $N\!F$ in higher dimensions.}
\label{fig:correlation_plot_comparison}
\end{figure}

By applying these 4 discretizations, we now investigate the time-bin correlations between Alice and Bob (Fig.~\ref{fig:correlation_plot_comparison}). We consider a noise level which is high enough to yield a negative witness for $d=10/20$ and a positive witness for $d=40/80$, certifying a separable and an entangled state respectively. The correlated events on the diagonal can be attributed to the maximally energy-time-entangled state emitted by our photon pair source, while the off-diagonal elements arise from our noise source. As a consequence of increasing dimensions, the noise is spread quadratically over the off-diagonal elements, while the correlated events on the diagonal spread linearly with the dimension, which is the key mechanism of Pathway I. However, discretizing to higher dimensions comes at the cost of additional noise induced by measuring close to the time resolution of the single photon detectors. This increasing crosstalk is clearly visible in Fig.~\hyperref[fig:correlation_plot_comparison]{\ref{fig:correlation_plot_comparison} (c)} and  \hyperref[fig:correlation_plot_comparison]{(d)}.

\subsection{Pathway I - Error analysis}
\label{sec:error}
In order to produce the error bars for the plot of entanglement detection in Fig.~3 of the main text, we ran a random number generator with Poisson distribution over the experimental data sets. Specifically, we assumed that the photon detections in the count matrices represented the Poissonian mean of the distribution. This comes with the tacit assumption that the probability of photon detection within a certain time interval does not change over the course of the experiment and that the probability of a detection in a particular time interval is independent of the probability of a detection in any other non-overlapping interval. We simulated $150$ new data sets, on which we computed an average witness $W(\rho_{\text{sim}})$ and the subsequent standard deviation.

\subsection{Pathway II - Correlation measurements}
\label{sec:correlation}

\begin{figure}[H]
\centering
\includegraphics[width=1\textwidth]{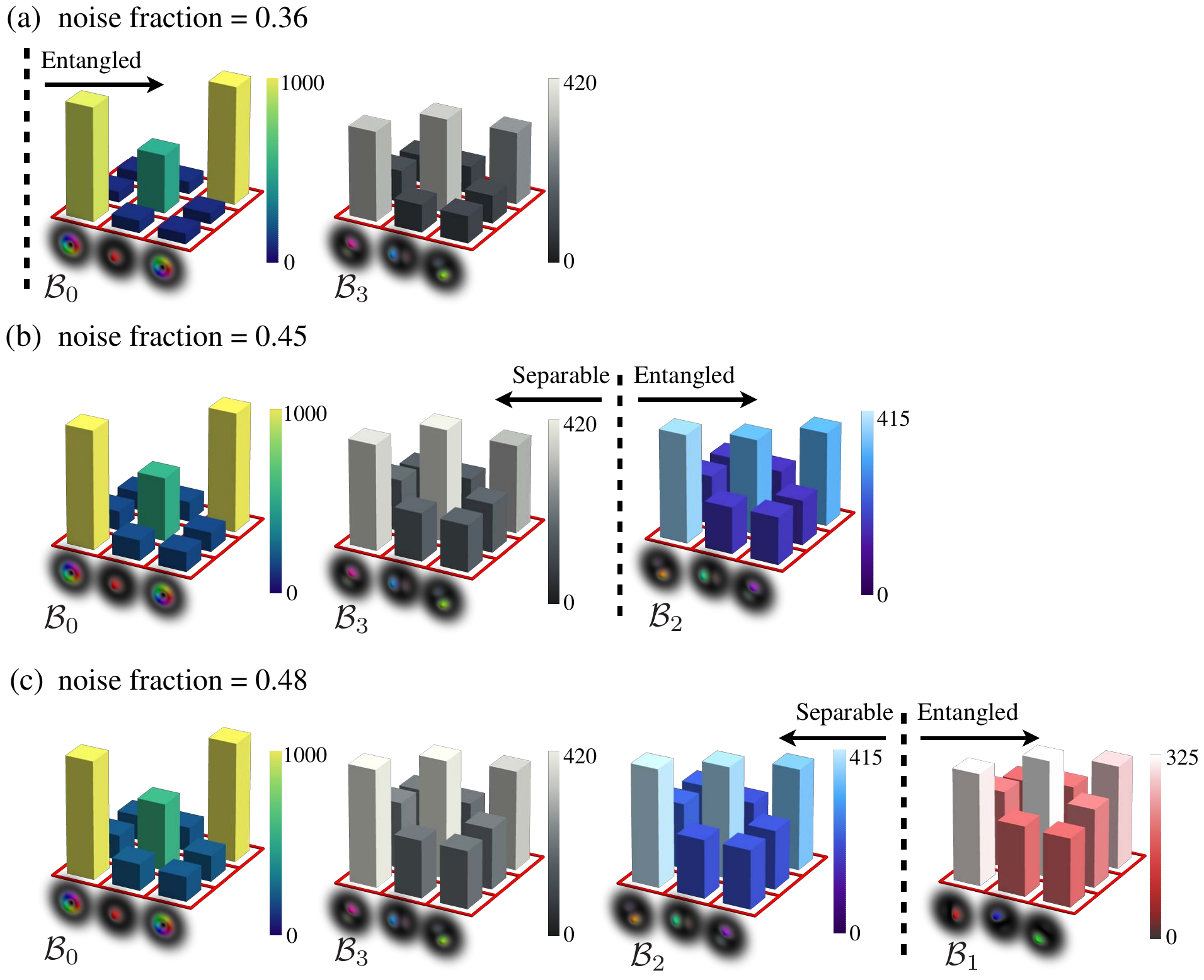} 
\caption{Detecting three-dimensional OAM entanglement between two photons with increasing noise fraction. Each graph shows the correlation measurements, given in coincidence counts per second, between Alice's (x-axis) and Bob's (y-axis) photon for all modes of a mutually unbiased basis (MUB). For a low noise fraction, the measurements in two MUBs are already enough to verify entanglement as depicted by the threshold (dashed line) on the left side in (a). When the noise fraction is increased the threshold moves to the right, which means that more MUBs need to be measured to still verify entanglement, as can be seen in (b) and (c). Only high-dimensionally entangled states allow to measure in more than 3 MUBs, which is the fundamental idea behind Pathway II to noise resilience.}
\label{fig:correlation_plot_OAM}
\end{figure}

Here, we present direct correlation measurements for the case of OAM entanglement with $d=3$. In Pathway II, larger noise fraction can be tolerated by including measurements in additional mutually unbiased bases (MUB). Correlation matrices, given in terms of coincidence counts, are shown for three different noise fractions, see Fig.~\ref{fig:correlation_plot_OAM}. In Fig.~\hyperref[fig:correlation_plot_OAM]{\ref{fig:correlation_plot_OAM}(a)}, we were able tolerate a noise fraction of 0.36 up to which entanglement can still be certified with measurements in only 2 MUBs. The measurements were done in MUB ${\cal B}_0$ (the OAM computational basis) and MUB ${\cal B}_3$, where the visibility sum ($\sum V$) is still larger than 4/3. We recall that the upper bound for separable states is given by $\sum_{j=0}^{k-1} V^{(j,j)}\leq 1+\frac{k-1}{d}$, where $k$ is the number of MUBs considered. In Fig.~~\hyperref[fig:correlation_plot_OAM]{\ref{fig:correlation_plot_OAM}(b)}, we considered the measurements in three MUBs, which led to a verification of entanglement up to a noise fraction of 0.45. Here, the measurements in MUBs ${\cal B}_0$, ${\cal B}_2$, and ${\cal B}_3$ were taken into account, where the visibility sum ($\sum V$) is still larger than 5/3. Finally, in Fig.~\hyperref[fig:correlation_plot_OAM]{\ref{fig:correlation_plot_OAM}(c)}, we performed measurements in all 4 MUBs to show the largest resilience to noise. For a noise fraction of up to 0.48, we were able to verify entanglement, where the visibility sum ($\sum V$) is still larger than 2.

\end{document}